# SUPERSYMMETRIC MODELS WITH A GAUGE SINGLET AND NON-UNIVERSAL SOFT TERMS FROM ORBIFOLD STRING THEORY

**Ph. Brax[1] @**

Department of Applied Mathematics and Theoretical Physics
Silver Street, Cambridge, CB3 9EW, England

**U. Ellwanger[2]**

LPTHE, Bât. 211, Université de Paris-Sud
F-91405 Orsay Cedex, France

**C.A. Savoy[3]**

Service de Physique Théorique, CE-Saclay
F-91191 Gif-sur-Yvette Cedex, France

ABSTRACT

The particle spectrum of the supersymmetric extension of the standard model with a gauge singlet is studied. Soft supersymmetry breaking terms are explictly chosen to be non-universal according to orbifold string theory . They depend on modular weights of chiral fields and on an angle $\theta$ specifying the supersymmetry breaking sector. Imposing radiative weak symmetry breaking and requiring that standard model Yukawa couplings should be allowed yield constraints on modular weights and almost specifies the angle $\theta$. We then perform a numerical analysis of the low energy spectrum. It turns out that the spectrum is very constrained, revealing salient features such as light Higgses and neutralinos. The latter turn out to be essentially gaugino-like.

---
[1] On Leave of Absence from SPhT, CE-Saclay, F-91191 Gif-sur-Yvette cedex, France
@ P.Brax@amtp.cam.ac.uk
[2] ellwange@qcd.th.u-psud.fr
[3] savoy@amoco.saclay.cea.fr

It is well-known that $SU(2) \times U(1)$ symmetry breaking in supersymmetric extensions of the standard model has to be radiatively induced through the quantum corrections to the soft supersymmetry breaking parameters of the scalar potential[1,2]. This phenomenon has been extensively discussed in the literature, mostly within the framework of the so-called minimal supersymmetric extension of the standard model (MSSM)[see, e.g., ref.3 for recent work]. There are also similar studies of an alternative model characterised by the addition of a chiral gauge singlet Y ((M+1)SSM)[4,5,6]. The dynamically determined vacuum expectation value (vev) $<Y>$ generates the effective coupling between the two Higgs doublets, solving the so-called "$\mu$-problem". A recent thorough analysis of this (M+1)SSM has established an interesting pattern of gauge symmetry breaking[6]. [1] Indeed, in this model, the singlet scalar vev $<Y>$ which turns out to be relatively large, is essentially induced as a tree level effect due to the supersymmetry breaking trilinear $Y^3$ coupling. Nevertheless there is a snag as non-trivial vacua emanating from large supersymmetry breaking trilinear terms (called A-terms thereafter) are well-known potential sources of color and electromagnetism breaking in supersymmetric extensions of the standard model. This imposes to set upper bounds to the quark and lepton A-terms[4,7].

A very popular assumption is that fermions and scalars are universally coupled to supergravity and to the unknown hidden sector inducing supersymmetry breaking. As a consequence, A-terms will reflect this flavour independence. This will be spoilt by radiative corrections. In the (M+1)SSM framework there seems to be a contradiction between this sort of universality and the requirement of a A-term in the singlet sector larger than A-terms in the light matter sector. However radiative corrections, in particular those due to the gauge interactions, nicely explains this paradox provided that the universal (tree level) parameter A fulfils some inequalities. Nevertheless the flavour independence of supergravity couplings is quite a strong an assumption and one could accordingly argue against its naturalness. Indeed this would require an extremely large symmetry at the Planck scale delicately broken by the observed gauge and Yukawa couplings.

The most valuable hints about the structure of the supergravity Lagrangian comes from superstring models[8]. As a rule soft terms are expected to be non-universal, for instance in models of orbifold compactification. Recently there have been several analyses of the MSSM with simple superstring-inspired parametrisations of the supersymmetry breaking effects[9,10]. In this paper we study the (M+1)SSM with non-universal supersymmetry breaking terms. In order to reduce the choice of the parameters we follow current investigations of orbifold compactification models. Within this context, a modular weight is assigned to each chiral multiplet. Then roughly speaking the contributions to the supersymmetry breaking terms split in two pieces; the dilaton sector is universal whereas the moduli sector is prportional to the modular weights, ie mostly non-universal[11]. Requiring the right pattern of $SU(2) \times U(1)$ breaking these parameters are severely restricted. Moreover the possible Yukawa couplings restrict the modular weight values. There are only a small number of allowed combinations. We then proceed to calculate the particle spectrum and impose experimental bounds. We follow closely ref [6].

Let us first introduce the superpotential:

$$W = \lambda Y H_1 H_2 + \frac{\kappa}{3} Y^3 + h_U H_1 U Q + h_D H_2 D Q + h_L H_2 E L \tag{1}$$

where $Y$ is a gauge singlet scalar, $H_1, H_2$ are the two Higgs doublets, $U, D, E$ are the scalars associated to the right-handed t-like, b-like and $\tau$-like fermions, and finally $Q$ and $L$ are the partners of the $SU(2)$ doublets. Here we assume that the top yukawa coupling $h_t$ is much larger than all the other quark and lepton Yukawa couplings. Introducing the usual notation $<H_2>=$

---

[1] More precisely the general pattern of the parameter space such that the sparticle spectrum is consistent with experimental bounds has been determined

$v\cos\beta$, $<H_1>=v\sin\beta$, we have to require that $\tan\beta < \frac{m_t}{m_b}$. The $\kappa$ and $\lambda$ parameters are free. Very large values of $\kappa$ and $\lambda$ lead to a wrong pattern of electroweak breaking.

As it stands the superpotential has an obvious $Z_3$ symmetry required to prevent any hierarchy problem. It also excludes supersymmetric mass terms. The soft terms are as follows:

$$-\mathcal{L}_{soft} = M_a\lambda_a\lambda_a + m_Y^2|Y|^2 + m_1^2|H_1|^2 + m_2^2|H_2|^2 + m_U^2|U|^2 + m_Q^2|Q|^2 + m_D^2|D|^2 + m_L^2|L|^2$$
$$+ m_E^2|E|^2 + A_\kappa\frac{\kappa}{3}Y^3 + A_\lambda\lambda Y H_1 H_2 + A_U h_U H_1 U Q + A_D h_D H_2 D Q + A_L h_L H_2 E L \quad (2)$$

where the $M_a$'s are Majorana masses of the gauginos. As above stated, non-universal parameters stem from orbifold compactification models. Yet the supersymmetry breaking mechanism remains mainly unknown in this framework. In order to proceed we follow the phenomenological approach of recent works[10]. Supersymmetry breaking is assumed to occur within the dilaton-moduli sectors thanks to non-vanishing vev of the auxiliary fields; $F_S$ for the dilaton and $F_{T_i}$ for the moduli. Furthermore we only retain the overall modulus $T$ which fixes the compactification scale and dismisses possible phases which could be sources of $CP$ violation. Define the 'goldstino angle' as[10]

$$\tan\theta = \frac{F_S}{F_T} \quad (3)$$

Then we can write the following soft terms at the compactification scale:

$$m_i^2 = \frac{M_0^2}{3}(1 + (n_i + 1)\cot^2\theta)$$
$$A_{ijk} = M_0(1 + (3 + n_i + n_j + n_k)\frac{\cot\theta}{\sqrt{3}}) \quad (4)$$
$$M_a = M_0$$

where $i, j, k$ denote chiral supermultiplets. The orbifold compactification of the heterotic superstring admits an $SL(2, Z)$[12] symmetry acting on the modulus $T$ and each matter supermultiplets $\Phi_i$. The modular weight of a supermultiplet specifies the way it transforms under this duality group:

$$T \to \frac{a + bT}{c + dT} \quad and \quad \Phi_i \to \Phi_i(c + dT)^{n_i} \quad (5)$$

where modular weights are integer $n_i = 0, -1, -2, -3$. This fixes the resulting soft terms. Several points should be stressed: i) In general the supersymmetry breaking parameters are matrices in the family space of quarks and leptons. For simplicity, we altogether neglect generation mixings so that supergravity couplings are assumed to be diagonal in the physical quark and lepton basis. ii) Although they have been derived in the framework of orbifold compactification, formulae (4) are to some extent valid for general superstring constructions.

Phenomenological constraints on the supersymmetry breaking terms of the (M+1)SSM become contraints on the modular weights and the goldstino angle. Let us first consider necessary conditions on the electroweak symmetry breaking minimum. We first recall that the $Y$ scalar develops a large classical value if the following condition is satisfied by the cubic and quadratic terms:

$$A_\kappa^2 > 9m_Y^2 \quad (6)$$

This condition (to be imposed at a scale of order $\frac{A_\kappa}{\kappa}$) is confirmed by numerical computations. Indeed, small $Y$ values are excluded once experimental bounds are imposed [6]. On the other hand, the evaluation at a lower scale than the compactification scale does not effectively affect the obtained bound as small values of $\kappa$ and $\lambda$ are favoured[6]. This entails the following condition:

$$\cot\theta > -\frac{1}{\sqrt{3}n_Y}(1+\sqrt{\frac{3n_Y+1}{n_Y+1}})$$
$$\cot\theta < -\frac{1}{\sqrt{3}n_Y}(1-\sqrt{\frac{3n_Y+1}{n_Y+1}}) \qquad (7)$$
$$n_Y < -1$$

The condition $n_Y < -1$ implies that the singlet $Y$ must be a twisted field in the orbifold compactification. The value $n_Y = -2$ turns out to be the unique solution entailling that either $\cot\theta > \frac{1+\sqrt{5}}{2\sqrt{3}}$ or $\cot\theta < \frac{1-\sqrt{5}}{2\sqrt{3}}$.

A second important condition ensures that color and electromagnetism are conserved at low enegy. As stated in the literature, the most dangerous charged vev are $<E>\sim<L>\sim<H_2>= O(\frac{A_L}{h_L})$, only significant for the first family. The necessary condition [2] is approximately given by the opposite condition to (6):

$$A_L^2 < 3(m_E^2 + m_L^2 + m_{H_1}^2) \qquad (8)$$

This is to be imposed at a scale $O(\frac{A_L}{h_L})$. Introduce the running parameter $t = \frac{1}{16\pi^2}\ln\frac{h_L\Lambda_C}{A_L}$ then the condition reads:

$$(1 - L_e + \frac{P}{\sqrt{3}}\cot\theta)^2 < 3(1 + K_e + \frac{P}{3}\cot^2\theta) \qquad (9)$$

where $P = 3 + n_E + n_L + n_{H_2}$ and the renormalised piece comes from the gauge couplings $L_e = 6t(g_1^2 + g_2^2)$ and $K_e = 12t(g_1^2(1 - 11g_1^2 t) + g_2^2(1 - g_2^2 t))$, finally $t \sim .13$, $g_1^2 \sim .13$ $g_2^2 \sim .43$. Let us notice that $P$ can only take the values $P = -2, -3$ (more about this later). This implies that bounds are easily deduced:

$$\begin{aligned} P = -2: & \quad \cot\theta > -1.07 \text{ or } < 1.44 \\ P = -3: & \quad \cot\theta > -.79 \text{ or } < 1.09 \end{aligned} \qquad (10)$$

Moreover experimental bounds on scalar masses further restrict the modular weights. The gauge radiative corrections are enough to ensure that squarks have relatively high masses whereas for slepton masses

$$m_E^2 \sim \frac{M_0^2}{3}(1 + (n_E + 1)\cot^2\theta) + \frac{M_0^2}{6} - \cos 2\beta \sin^2\theta_W M_Z^2$$
$$m_L^2 \sim \frac{M_0^2}{3}(1 + (n_L + 1)\cot^2\theta) + \frac{M_0^2}{2} + \cos 2\beta \frac{M_Z^2}{2} \qquad (11)$$

a sufficient condition is

$$n_E > -3, \; n_L > -3 \qquad (12)$$

These considerations give already interesting bounds on the goldstino angle and some modular weights. The latter have been considered as free parameters. However orbifold compactifications impose stringent constraints on these integers (see further). Hence we take into account this small number of possible weights in our numerical study. Our analysis hinges on the hypothesis that all Yukawa couplings in (1) (only the third family is involved) are trilinear terms at the compactification level. In view of the smallness of some Yukawa couplings, this could be challenged and some of them could stem from non-renormalizable interactions. This would highly clutter our analysis and make our results depend on unknown fields and their modular weights. To be concrete we choose on the contrary to stick to the standard approach and assume that all those Yukawa couplings are trilinear terms at the compactification scale. Then we simply repeat the third family pattern of modular weights for the first and second families of quarks and leptons so that FCNC effects are maximally diminished. Following this trend we are able to derive conspicuous results about the allowed modular weights. They are given in table II . Notice that $n_Y = -2$ and $P = -2, -3$ are obtained. The following part gives a detailled calculation of modular weights, the numerical analysis and its phenomenological consequences are then resumed.

---

[2] As stated in ref.[4], this condition applies only if the relevant Yukawa couplings are small with respect to gauge couplings and the relevant scalar masses are not very different. This is clearly fulfilled in the present model.

The modular weights of the supersymmetric standard model sparticles can be tightly restricted thanks to a close relationship with yukawa couplings. Indeed for the types of (0,2) orbifolds here considered, ie symmetric coxeter orbifolds, stringent selection rules are to be imposed on Yukawa Couplings. They stem from the following requirements:

i) Space and point group invariance: each field is associated with a conjugacy class of the point group, a couple $g = (\theta^j, (1 - \theta^k)(\theta^k f + \Lambda))$, $k = 0..j - 1$ where $\theta^j$ specifies the twisted sector and $f$ is a fixed point of $\theta^j$. The point group is the semi-direct product of the space group generated by $\theta$ and the weight lattice $\Lambda$. Yukawa Couplings $\phi_1 \phi_2 \phi_3$ are allowed provided the product of the associated group elements $g_1 g_2 g_3$ contains the identity. This condition implies the point group selection rule :

$$j_1 + j_2 + j_3 \equiv 0 \ (\mathrm{mod} N) \tag{13}$$

where $N$ is the order of $G$.

ii) Invariance under plane rotations: each plane of the internal six dimensional lattice can be independently rotated; this is a symmetry of the interactions which restricts the number of derivatives $\partial X$. For the case at stake this implies that Yukawa couplings involving left oscillating twisted states $\prod_{i,j=1}^{3} (\alpha_{m_i + \theta_i}^i)^{p^i} (\alpha_{n_j - \theta_j}^j)^{q^j} |\sigma> \otimes |P + \delta>$ , where $|\sigma>$ is a twisted state and $|P + \delta>$ is a state of the $E_8 \times E_8$ sector, have to satisfy:

$$\sum_{m=1}^{3} (p_m^i - q_m^i) \equiv 0 \ (\mathrm{mod} N) \tag{14}$$

This selection rule is directly linked to the modular weights of particles.

The modular weight of a field, if the overall modulus is only taken into account, depends on the number of left and right oscillators $p = \sum_{i=1}^{3} p^i$, $q = \sum_{i=1}^{3} q^i$. Three cases are to be envisaged: $n = -1$ (untwisted), $n = -2 - p + q$ (twisted, all planes rotated), $n = -1 - p + q$ (twisted, one plane unrotated). Notice that for twisted fields, modular weights differ by one unit depending on whether their twist $\theta^j$ leaves one plane unrotated. In the sequel we will only deal with matter fields in a twisted sector.

The available range of modular weights is further restricted as fields arise from the massless spectrum of the heterotic string theory. The left mass formula depends on the fractional number of oscillators:

$$\frac{1}{8} M_L^2 = N_{osc} + h_{KM} + E_0 - 1 \tag{15}$$

where $E_0 = \frac{1}{2} \sum_{i=1}^{3} |v_i|(1 - |v_i|)$ is the vacuum energy and $(v_i)$ is the twist vector ($\sum_i v_i = 0$) corresponding to a rotation by an angle $2\pi v_i$ of the ith plane. The last term $h_{KM}$ corresponds to the conformal dimension of the matter fields from the left-moving $E_8 \times E_8$ part. As the gauge group contains the standard model $SU(3) \times SU(2) \times U(1)$ the conformal dimension is a sum of contributions from each group $G_\alpha$; $h_{KM} = \sum_\alpha \frac{C(R_\alpha)}{C(G_\alpha)+k_\alpha}$; where $C(R)$ is the Casimir of the representation $R$ and $C(G)$ the Casimir of the adjoint representation. We will restrict ourselves to the grand unified values of the Kac-Moody levels $k$ where $k = 1$ for non-abelian groups and $k = \frac{5}{3}$ for the hypercharge. The conformal dimension is always greater than its value for the standard model gauge group. In that case it reads $k = \frac{3}{5}$ for sparticules of types $Q, U, E$ and $k = \frac{2}{5}$ for $L, H, D$. Notice that no contribution appears for the singlet field $Y$. The allowed values for modular weights are summed up in the following table:

| $Z_6$ | $Z_{12}$ | $(\theta_1, \theta_2, \theta_3)$ | Y | Q,U,E | L,H,D |
|---|---|---|---|---|---|
| $\theta^2$ | $\theta^4, \theta^8$ | $(\frac{1}{3}, \frac{1}{3}, \frac{1}{3})$ | $-4 \leq n \leq -1$ | $n = -2$ | n=-2 |
| $\theta$ | $\theta^2, \theta^{10}$ | $(\frac{1}{3}, \frac{1}{6}, \frac{1}{6})$ | no | $n = -2$ | $-4 \leq n \leq -1$ |
| no | $\theta, \theta^5, \theta^7, \theta^{11}$ | $(\frac{1}{3}, \frac{1}{12}, \frac{5}{12})$ | no | $-3 \leq n \leq -2$ | $-5 \leq n \leq -2$ |
| $\theta^3$ | $\theta^6$ | $(0, \frac{1}{2}, \frac{1}{2})$ | no | $n = -1$ | $n = -1$ |
| no | $\theta^3, \theta^9$ | $(0, \frac{1}{4}, \frac{1}{4})$ | no | $n = -1$ | $-2 \leq n \leq 0$ |

Table 1: Allowed modular weights for massless fields

Combining the above-mentioned results yields the possible modular weights for the supersymmetric standard model field in the singlet model. First of all the existence of a self coupling $Y^3$ implies that the order of the orbifold is either $N = 6, 12$. This is a consequence of (13). The $Z_6$ orbifold whose twist vector is $\theta = (\frac{1}{6}, \frac{1}{6}, \frac{2}{3})$ allows the self coupling $\theta^2 \theta^2 \theta^2$. On the other hand the $Z_{12}$ orbifold whose twist vector is $\theta = (\frac{1}{12}, \frac{1}{3}, \frac{7}{12})$ allows a self coupling $\theta^4 \theta^4 \theta^4$. We are now ready to check whether the $H_1 H_2 Y, H_1 Q U, H_2 Q D, H_2 L E$ couplings are allowed. Let us begin with the $Z_6$ case as it is more convenient. The $Z_{12}$ case can be treated in the same vein. Let us examine the oscillator rules for the $Y^3$ coupling. As $Y$ is assigned to the $\theta^2$ sector whose twist is $(\frac{1}{3}, \frac{1}{3}, \frac{1}{3})$ the allowed range is $n_Y \in \{-4, -3, -2, -1\}$. The value $-1$ is forbidden by the minimum condition. Moreover the oscillator rule implies that $3n_Y \equiv 0 \ (6)$ which imposes that $n_Y \in \{-4, -2\}$. The only allowed coupling for this $Z_6$ orbifold are $\theta^2 \theta^2 \theta^2$ and $\theta \theta^2 \theta^3$. This entails that $H_1$ and $H_2$ can either be twisted as $\theta, \theta^3$ (respectively $\theta^3 \ \theta$) or $\theta^2, \theta^2$.

Suppose that the latter is true, this signifies that their twist is $\theta^2 = (\frac{1}{3}, \frac{1}{3}, \frac{1}{3})$. Consequently the modular weights of the Higgs field and singlet field are uniquely determined: $n_{H_1} = n_{H_2} = n_Y = -2$. The twisted sector of Q fields can either be $\theta, \theta^2, \theta^3$ entailing that U and D fields lie in the same twisted sector respectively $\theta^3, \theta^2, \theta$. Similarly E and L fields can be assigned to the same arrangement of twists as Q and D fields. Two cases arise, either all fields in a Yukawa coupling rotate all three planes, ie $\theta^2 \theta^2 \theta^2$, or there is a field whose twist leaves one plane unrotated, ie $\theta^3 = (\frac{1}{2}, \frac{1}{2}, 0)$ for the $\theta \theta^2 \theta^3$ coupling. The oscillator rule for the former imposes that the sum of modular weights is $n_1 + n_2 + n_3 \equiv 0 \ (6)$ whereas for the latter $n_1 + n_2 + n_3 \equiv -5 \ (6)$. Complying with these equalities requires that the following pairing between twists and modular weights is achieved $\theta \rightarrow -2$, $\theta^2 \rightarrow -2$, $\theta^3 \rightarrow -1$.

Suppose now that the Higgs fields are assigned to the following twists $H_1 \rightarrow \theta$, $H_2 \rightarrow \theta^3$.

The oscillator rule imposes that the modular weights of the Higgs and singlet fields are $n_{H_1} = n_Y = -2$, $n_{H_2} = -1$. Furthermore the Q fields have to be twisted as $\theta^2$ entailing that U fields are twisted as $\theta^3$ and D fields as $\theta$. The oscillator rule reads for these couplings involving a twist leaving one plane unrotated is $n_1 + n_2 + n_3 \equiv -5$ (6). Therefore the squarks have modular weights $n_U = -1$, $n_D = -2$. Finally the sleptons can either be twisted as $\theta^2$ and $\theta$ yielding modular weights $n_E = -2; n_L = -2$.

Finally let us examine the last assignement $H_1 \to \theta^3$, $H_2 \to \theta^2$. In that case the modular weights of the higgs and singlet fields are exchanged $n_{H_2} = n_S = -2$, $n_{H_1} = -1$. The Q fields are still twisted as $\theta^2$ and their modular weights remain $n_Q = -2$. The squarks U and D are now twisted as $\theta$ and $\theta^3$ enforcing their modular weights to be $n_U = -2$, $n_D = -1$. The sleptonic sector can be twisted as $\theta^3$ and $\theta^2$ increasing the number of modular weights to $n_L = -1, -2; n_E = -2, -1$. All these combinations are summed up in the table 2.Notice that the singlet modular weight is unique. The squark and slepton sectors are disentangled, their modular weights can be varied independently. It is noteworthy that no oscillators are allowed, fields are twisted but never oscillating.

The same analysis as above can be performed for the $Z_{12}$ orbifold. Unfortunately it is much more tedious. The end result is confined in tables 2 and 3.

| Y | $H_1$ | $H_2$ | D | Q | U | E | L |
|---|---|---|---|---|---|---|---|
| -2 | -2 | -2 | -1 | -2 | -1 | -2 | -2 |
| -2 | -2 | -2 | -1 | -2 | -1 | -2 | -1 |
| -2 | -2 | -2 | -1 | -2 | -1 | -1 | -2 |
| -2 | -2 | -2 | -2 | -2 | -2 | -2 | -2 |
| -2 | -2 | -2 | -2 | -2 | -2 | -2 | -1 |
| -2 | -2 | -2 | -2 | -2 | -2 | -1 | -2 |
| -2 | -2 | -2 | -2 | -1 | -2 | -1 | -2 |
| -2 | -2 | -2 | -2 | -1 | -2 | -2 | -1 |
| -2 | -2 | -2 | -2 | -1 | -2 | -2 | -2 |
| -2 | -1 | -2 | -1 | -2 | -2 | -1 | -2 |
| -2 | -1 | -2 | -1 | -2 | -2 | -2 | -1 |
| -2 | -2 | -1 | -2 | -2 | -1 | -2 | -2 |

Table 2: $Z_6$-Modular weights of sparticles

| Y | $H_1$ | $H_2$ | D | Q | U | E | L |
|---|---|---|---|---|---|---|---|
| -2 | -2 | -2 | -2 | -2 | -2 | -2 | -2 |
| -2 | -2 | -2 | -2 | -2 | -2 | -2 | -1 |
| -2 | -2 | -2 | -2 | -2 | -2 | -1 | -2 |
| -2 | -2 | -1 | -2 | -2 | -2 | -2 | -2 |
| -2 | -1 | -2 | -1 | -2 | -2 | -1 | -2 |
| -2 | -1 | -2 | -1 | -2 | -2 | -2 | -1 |
| -2 | -1 | -2 | -1 | -2 | -2 | -3 | -0 |
| -2 | -2 | -2 | -2 | -1 | -2 | -1 | -2 |
| -2 | -2 | -2 | -2 | -1 | -2 | -2 | -1 |
| -2 | -2 | -2 | -2 | -1 | -2 | -2 | -2 |
| -2 | -2 | -2 | -1 | -2 | -1 | -2 | -1 |
| -2 | -2 | -2 | -1 | -2 | -1 | -1 | -2 |
| -2 | -2 | -2 | -1 | -2 | -1 | -2 | -2 |

Table 3: $Z_{12}$-Modular weights of sparticles

Notice that as for the $Z_6$ orbifold the singlet field has an unique modular weight $n_S = -2$. The squarks can have modular weights -1 and -2 whereas the range is from -3 to 0 for sleptons. The modular weights of the standard model sparticles are tightly correlated. This implies strong correlations between the sparticle masses at the compactification scale. As a matter of fact the low energy spectrum of sparticles is very constrained.Notice that the modular weight $n_E = -3$ is forbidden by the slepton mass condition.

We proceed to discuss the phenomenological implications of the (M+1)SSM.For this purpose we scan the parameter space for each set of modular weights, ie we vary $\lambda, \kappa, h_t, \tan\theta$ while the overall supersymmetry breaking scale $M_0$ is fixed by the physical value of the Z-boson mass. We have imposed experimental bounds on the masses of the sparticles. Bounds on the top mass are also included. As the neutralino mass matrix is diagonalised we have ckecked that the lightest neutralino is always lighter than the lightest chargino. The bound to the neutralino, neutral scalars and sleptons contributions to the $Z_0$ width is also taken into account. We have examined $O(10^6)$ combinations of parameters for both types of models, ie $P = -2, -3$. The impact of radiative corrections to the scalar potential on the resulting spectrum is noteworthy. This has been taken into account as discussed in ref.[6]. We none the less present a rough analysis of the minimum conditions neglecting radiative corrections which lead to interesting connections between parameters. It has been established in ref.[6] (and our numerical calculations are in very good agreement with it) that the phenomenologically acceptable (M+1)SSM vacua follow a rather simple pattern. The $<Y>$ vev is basically fixed by the purely $Y$ sector and is approximately given by:

$$\nu = \kappa <Y> = -\frac{A_\kappa}{4}(1 + \sqrt{1 - \frac{8m_Y^2}{A_\kappa^2}}) \qquad (16)$$

Since from previous considerations $p_Y = n_y + 1 = -1$ and $\cot\theta \sim 1$, $\frac{m_Y^2}{A_\kappa^2}$ tends to be relatively small then $\nu \sim M_0 \frac{(1-\sqrt{3}\cot\theta)}{2}$ provides a reasonable approximation. Numerically we always find $<Y> \geq 1 TeV$. Then the $SU(2) \times U(1)$ breaking vev's nearly follow the corresponding conditions on the MSSM with the effective parameters[6]:

$$\mu = \frac{\lambda\nu}{\kappa}$$
$$B = A_\lambda + \nu \qquad (17)$$

The low energy soft terms can be written in terms of their values at the compactification scale (4) using solutions of the corresponding renormalisation group equations. In the case $\frac{h_t}{\lambda} > 1$ which turns out to be almost always true, relatively simple expressions are given in ref.[6]. The relevant ones are the following:

$$A_\lambda = A_{\lambda 0} - \frac{\rho}{2}A_{t0} + (\frac{1}{2} - \rho)M_0$$
$$m_2^2 = m_{20}^2 + 0.5 M_0^2 \qquad (18)$$
$$m_1^2 = m_{10}^2(1 - \frac{\rho}{2}) - \frac{\rho}{2}(m_{t0}^2 + m_{q0}^2) - \frac{\rho}{2}(1-\rho)(A_{t0} + 2M_0)^2 - (2\rho - 0.5)M_0^2$$

where $\rho = \frac{h_t^2}{h_{t0}^2} \sim 0.82 h_t^2$. Indices "0" refer to the soft terms at the compactification scale (4) and the dependence on the gauge couplings has been numerically given. Then then parameters in (17,18) are approximately given by:

$$\mu = \frac{\lambda M_0}{2\kappa}(1 - \sqrt{3}\cot\theta)$$
$$\frac{B}{M_0} \sim -0.5\rho - \frac{\cot\theta}{\sqrt{3}}(P + \frac{3}{2} - \frac{\rho}{2}P')$$
$$\frac{m_2^2}{M_0^2} \sim \frac{1}{3}(\frac{5}{2} + (n_2 + 1)\cot^2\theta) \qquad (19)$$
$$\frac{m_1^2}{M_0^2} \sim \frac{1}{3}(\frac{5}{2} + (n_1 + 1)\cot^2\theta) - 3\rho + \frac{\rho}{2}(\rho - \frac{P'\cot^2\theta}{3} + (1-\rho)\frac{P'\cot\theta}{\sqrt{3}})$$

where $P' = 3 + n_1 + n_U + n_Q$. From the well-known MSSM algebraic expresions for $\tan\beta$ and $M_Z^2$ in terms of $B, \mu, m_1^2, m_2^2$ one can study the approximate constraints on the parameters. Since $\tan^2\beta > 4$ for the obtained solutions, parameters can be roughly evaluated as follows; first notice that for $m_0 > M_Z$ one obtains $\mu^2 \sim -m_1^2$ so that :

$$|\frac{\lambda}{\kappa}| \sim \frac{2\sqrt{3}\rho}{|1 - \sqrt{3}\cot\theta|} \tag{20}$$

On the other hand, for the values of $\tan\beta$ (see fig.1) [3]:

$$\tan\beta \sim \frac{\sqrt{m_2^2 - m_1^2}}{|B|} \tag{21}$$

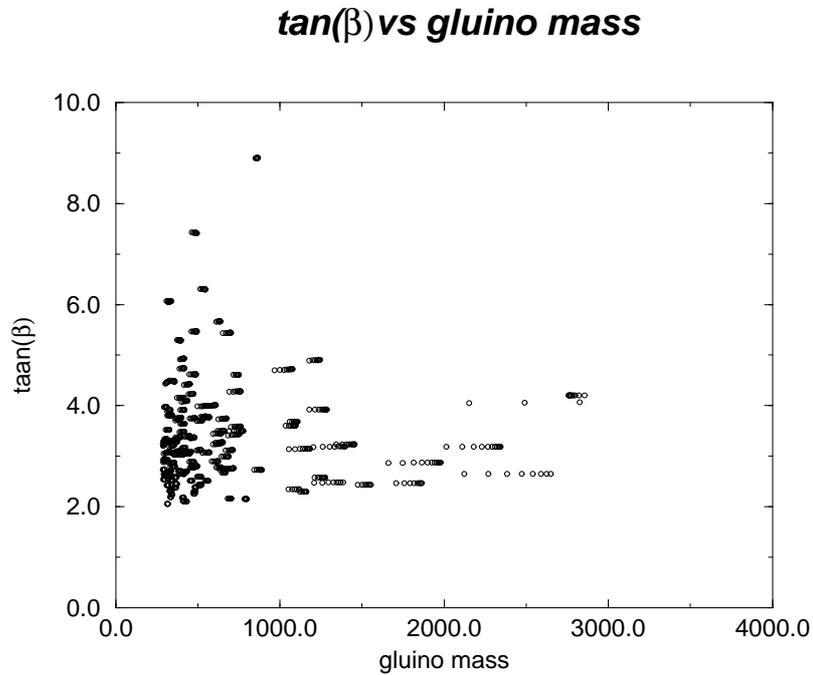

Fig.1:$\tan\beta$ as a function of $M_0$ when $\cot\theta > 0$

In order to obtain large values of gaugino masses one has to fine tune parameters, eg the ratio $\frac{\lambda}{\kappa}$. Nevertheless in our scanning of the parameter space we still find a reasonable amount of solutions with $M_0 \sim 1 TeV$.

Let us examine the numerical results. the following describes two typical situations. The case $P = -3$ is universal in the Higgs sector. It is a particular case of the parametrisation of ref[6], apart from slight variations of the compactification scale squark and slepton masses. Moreover fewer solutions are found in the allowed slot for $\cot\theta$ as it is very narrow.

---

[3] However $\tan\beta$ is very sensitive to radiative corrections

Let us now focus on the $P = -2$ case. The allowed range for $\cot\theta$ is broader so more solutions are found. We can separate positive and negative values of $\cot\theta$. Numerical results used for the figures are obtained with the modular weights corresponding to the tenth line of table 2. The values of $\cot\theta$ are shown in fig.2 as a function of the gluino mass ($\sim 2.7 M_0$).

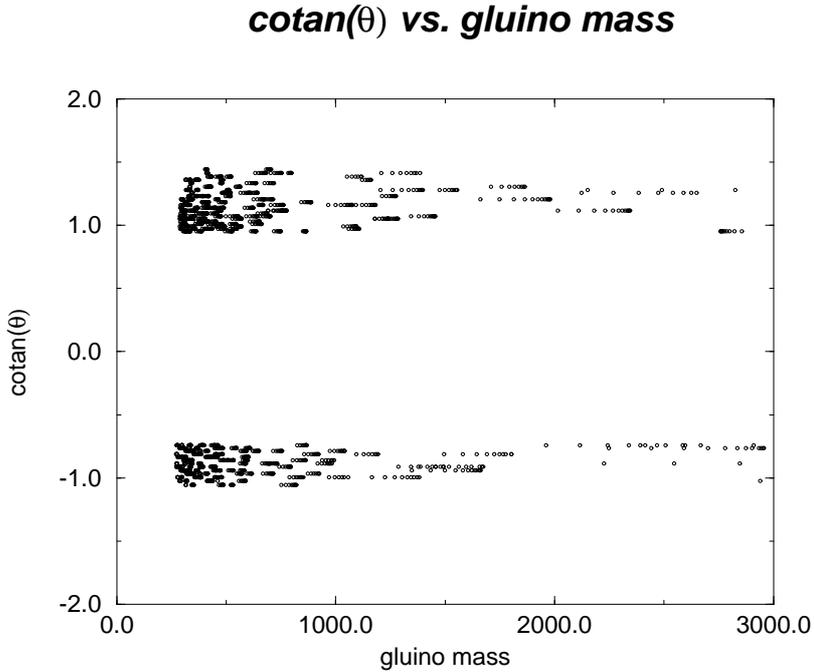

Fig.2: $\cot\theta$ as a function of the gluino mass

They are in agreement with the bounds (7) and (10) (but (10) is far from being saturated when $\cot\theta < 0$). The results for $\tan\beta$ are shown in Fig.1 for $\cot\theta > 0$ For negative $\cot\theta$, $\tan\beta$ has an opposite sign and tends to be larger by a factor $\sim 3$. It is worth noticing that one sign is strongly favoured with universal soft terms while in the non universal $P = -2$ case both signs are easily obtained by flipping the $\cot\theta$ sign. The Yukawa couplings are in the range $2.10^{-6} \leq \lambda_0^2 \leq .1$, $.1 \leq \frac{\lambda^2}{\kappa^2} < 3$ in qualitative agreement with (20). These ranges are consistent with the values of quark Yukawa couplings at the compactification scale. The mass spectrum of squarks and sleptons is almost a linear function of $M_0$ but sneutrinos can be as light as the present experimental bound. Gluino masses start at 250 GeV (while we only impose the experimental bound of 110 GeV). The chargino can be as light as the LEP1 limit. The neutralino is essentially gaugino-like and purely bino above 80 GeV. Below this, it is a bit more parameter dependent and can be as light as 20 GeV (predominantly photino). The lightest Higgs masses are depicted in Fig.3.

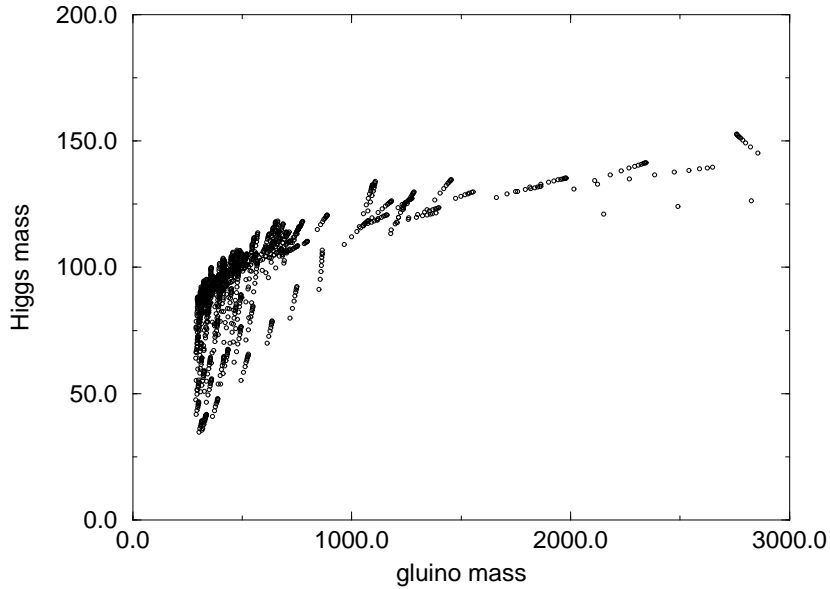

Fig.3: The Higgs mass as a function of the gluino mass

The preferred region is around 100 GeV but the Higgs can be as light as 35 GeV with a correspondingly weaker coupling to the Z-boson.

We have been able to derive strong correlations between particle masses in the singlet extension of the supersymmetric standard model. These constraints come from the particular form of the soft terms at the compactification scale of orbifold string theories. We should emphasize the possibility of very low masses in the sparticle spectrum, ie light neutralinos. The mildest breaking of universality we have herein analysed already shows that strong prediction at the weak scale can be deduced from physics at the Planck scale. The existence of a singlet vev of O(1) TeV gives restrictions on the possible modular weights of chiral weights and select some special patterns of soft terms at the compactification scale. We find that universal soft terms are excluded within this class of models; low energy physics requires a goldstino angle $\theta \sim \pm\frac{\pi}{4}$. This is nice interplay between string theory and low energy physics.